\documentclass[onecolumn,showpacs]{revtex4}
\usepackage{graphicx}% Include figure files
\usepackage{epstopdf}
\usepackage{dcolumn}% Align table columns on decimal point
\usepackage{bm}% bold math
\usepackage{amsmath}

\begin{document}
{\setlength{\oddsidemargin}{1.2in}
\setlength{\evensidemargin}{1.2in} } \baselineskip 0.55cm
\begin{center}
{\LARGE {\bf Evolution of Kaluza-Klein Like Wet Dark Fluid in $f(R,T)$ Theory of Gravitation}}
\end{center}
\date{\today}
\begin{center}
 Koijam Manihar Singh $^{2}$, S. Surendra Singh$^{1}$ and  Leishingam Kumrah$^{1}$\\

   1. School of Engineering and Technology, Mizoram University,Aizawl, P.O. Box No. 190.,India\\
   2. Department of Mathematics, NIT Manipur,\\ Imphal-795004,India\\
   Email:{ssuren.mu@gmail.com, drmanihar@rediffmail.com}\\
 \end{center}

\textbf{Abstract:}
Here we study the essence of $f(R,T)$ gravitation theory in five dimensional Universe and see the role of dark energy in the form of wet dark fluid in such a Universe. It is found that the dark energy is not exaggerated in contributing to the accelerating expansion of the Universe though the expansion is inherent as a result of the theory itself and due to the geometric contribution of matter. It is interesting to see that in some model it is found that there was some era before the beginning of the present era, and some of the model Universe came out to be either oscillatory or cyclic. Some of the models are seen to go to $\Lambda CDM$ models in late future as in Einstein gravitation theory, starting the evolution with a big bang. Most of the models undergo early inflation as well as late time accelerating expansion thus defining as good models for real astrophysical situations, with dark energy playing fundamental role in these Universe.


\textbf{Introduction:}
General theory of relativity can explain most of gravitational phenomena known to date. However the recent observation of accelerated expansion of the Universe could not be explained by the general theory of relativity. To accommodate the accelerated expansion of Universe, there has been considerable interest in deriving cosmological models
for alternative theory of gravitation or modified theories of general relativity and dark energy(Perlmutter, et al. 1998, Riess, et al. 1998). Many theories are developed subject to the known law of physics but still the driving force behind the accelerating Universe is a mystery . The accelerated expansion of the Universe strongly indicates the presence of a strong negative pressure called dark energy. However the nature of dark energy remains a mystery(Ratra and Peebles, 1988).  Different theoretical models of Universe have been proposed to explain the nature of dark energy and accelerated expansion of the Universe. Some of these models are quintessence(Ratra and Peebles, 1988), phantom energy(Carroll,2003), k-essence(Steinhardt et al. 2001), tachyons, f-essence and Chaplygin gas. As another approach, different modified theories of gravity such as $f(R)$(Carrol et al.2004,Capozziello, et al.2006,Nojiri and Odinsov 2006, Sharif and Arif 2012), $f(T)$(Bamba, et al. 2012, Jamil, et al. 2012, Momeni et al.2012)and Gauss-Bonnet gravity of $f(G)$(Garcia et al. 2011 and Nojiri et al. 2008) gravity has been formulated to study the nature of dark energy and accelerated expansion of Universe.\\

To study the nature of dark energy and late-time accelerating expansion of the Universe, we start from a theory defined by the action and solve the equation of motion to define the background dynamics. The cosmological reconstruction of gravitational theorem is considered where complicated background cosmology can be reconstructed which complies with the observed data. This eventually leads to a particular choice of the arbitrary function or potential, providing the qualitative difference between general relativity and modified gravity. Capozziello et al. (2006) and Nojiri and Odinstov (2006) have developed the general approach to reconstruction modified gravity and dark energy models.\\

Keeping this in mind, modified theory called $f(R,T)$ theory have been formulated by Harko et al. (2011) where the gravitational lagrangian $L_{m}$ is given by an arbitrary function of the Ricci scalar $R$, and trace of the stress-energy tensor $T$. Gravitational field equations in the metric form as well as the equations of motions for test particle obtained from the result of covariant divergence of the stress-energy tensor w.r.t the metric. The possibility of the reconstruction of Friedmann-Robertson-Walker cosmology by an appropriate choice of function $f(T)$ have been demonstrated by them.\\

Paul et al. (2009) studied the Friedmann-Robertson-Walker model in $f(R)$ gravity while Sharif and Shamir (2009,2010) investigated the solution of Bianchi type I and V space-time $f(R)$ gravity. Shamir(2010) studied the exact vaccum solution of Bianchi types-I and III and kantowski-Sachs space-time in the metric version of $f(R)$ gravity. Moreover dark energy models in modified theories of gravitation have been studied and investigated by several authors. Shamir and Bhatti (2012) study Bianchi type-III dark energy model in Brans and Dicke (1961) theory. Bianchi type-I dark energy model in scale covariant theory of gravitation has been discussed by Reddy et al. (2012a), while Bianchi type- II, III and V dark energy models in Saez and Ballester (1986) scalar theory of gravitation respectively have been investigated by Naidu et al. (2012a, 2012b). Rao and Nulima (2013) have discussed LRS Bianchi type-I dark energy cosmological model in general scalar tensor theory of gravitation. Bianchi type-II, VIII and IX perfect fluid dark energy cosmological model in Saez-Ballester and general theory of gravitation have been investigated by Rao et al. (2013). Reddy et al. (2012c) presented a five dimensional Kaluza-Klein dark energy model in Saez-Ballester theory. Reddy et al. (2012b) discussed Kaluza-Klein dark energy model in f(R,T) gravity while Sahoo and Mishra (2014) investigated and discussed Kaluza-Klein dark energy model in the presence of wet dark fluid in $f(R,T)$ gravity. Bianchi type-III dark energy model in $f(R,T)$ gravity have been obtained by Reddy et al.  (2013) and Bianchi type-II, VIII and IX perfect fluid cosmological models in this modified theory of gravity have been discussed by Rao et al. (2014). Recently, Rao and Suryanarayarana (2014) have discussed and investigated higher dimensional perfect fluid cosmological models in this theory.\\

In this paper, we focus our attention on exploring the solution of five dimensional Kaluza-Klien cosmological model in the content of $f(R,T)$ gravity. The present paper is organised as follows; Sect. 1, a brief introduction is given. Sect. 2 consist of a brief discussion about wet dark fluid as a candidate for dark energy. $f(R,T)$ gravity which is also a modification of general relativity is also discussed considering the case $f(R,T)= R+2f(T)$. In Sect. 3 gravitational field equation in $f(R,T)$ gravity is established with the help of Kaluza-Klien metric in the presence of wet dark fluid  (WDF) and four different cases are discussed. In Sect. 4 Solution obtained is presented and conclusions are given.\\

 2. {\bf  Field Equations and their solutions:} \setcounter {equation}{0}
\renewcommand {\theequation}{\arabic{equation}}\\
For our problem here we take wet dark fluid as a candidate for dark energy, and in the line of generalised Chaplygin gas its equation of state is taken in the form
\begin{equation}
  p_{WD}=(\rho_{WD}-A)\omega
\end{equation}
where $p_{WD}$ and $\rho_{WD}$ are respectively the fluid pressure and the energy density of wet dark fluid. It may be noted here that equation (1) can represent many fluids including water. $\omega$ and $\rho_{WD}$ are parameters which are taken to be positive with the condition
\begin{equation}
      0\leq \omega \leq 1
\end{equation}
      The parameter $\omega$ is such that if we take $c_{s}$ as the adiabatic sound speed in wet dark energy fluid, then
\begin{equation}
 \omega=c^{2}_{s}
\end{equation}
The energy density of the wet dark fluid can be obtained from the conservation equation
\begin{equation}
    \dot{\rho}_{WD}+ 3H(p_{WD}+\rho_{WD})=0
\end{equation}
Now using the relation
\begin{equation}
    3H=\frac{\dot{V}}{V}
\end{equation}
where $H$ is the mean Hubble constant and $V$ is the spatial volume of the Universe, in equation (4) and making use of equation (1) we obtain
\begin{equation}
\rho_{WD}=\frac{\omega}{1+\omega}A+\frac{c}{V^{(1+\omega)}}
\end{equation}
where $c$ is a constant of integration.\\
It may be noted that wet dark fluid comprises of two parts, one part behaving as a cosmological constant and another part behaving as a standard fluid with the equation of state
\begin{equation}
p=\omega \rho
\end{equation}
Now if $c>0$, then for this fluid the strong energy condition

    \begin{equation}
    p+\rho\geq0
    \end{equation}
   will not be violated. Hence we have
   \begin{equation}
    p_{WD}+\rho_{WD}=(1+ \omega)\rho_{WD}-\omega A=c(1+\omega)V^{-(\omega+1)}\geq 0
   \end{equation}
 Here for obtaining the fluid equation from the $f(R,T)$ gravitational theory, we use the Hilbert-Einstein type variational principle for which the action is taken as
   \begin{equation}
    S=\frac{1}{16\pi}\int[f(R,T)+L_{m}]\sqrt{-g}d^{4}x
   \end{equation}
where $L_{m}$ is the Lagrangian matter density.\\
Here the stress energy tensor of matter $T_{ij}$ is taken in the form

\begin{equation}
   T_{ij}=-2(-g)^{-\frac{1}{2}}\frac{\delta(\sqrt{-g}L_{m})}{\delta g^{ij}}
   \end{equation}
   and thereby the trace becomes
    \begin{equation}
   T=g^{ij}T_{ij}
   \end{equation}
  On the assumption that the matter Lagrangian depend solely on $g_{ij}$ rather than its derivatives we see that
    \begin{equation}
    T^{ij}=g^{ij}L_{m}-\frac{\partial L_{m}}{\partial g^{ij}}
   \end{equation}
  Then the field equation of $f(R,T)$ gravitation theory came out, by varying the action $S$ with respect metric tensor $g_{ij}$ to be
    \begin{equation}
    f_{R}(R,T)R_{ij}-\frac{1}{2}f(R,T)g_{ij}+(g_{ij}\nabla_{k}\nabla^{k}-\nabla_{i}\nabla_{j})f_{R}(R,T)=8\pi T_{ij}-f_{T}(R,T)T_{ij}-f_{T}(R,T)Q_{ij}
   \end{equation}
  Here
    \begin{equation}
    Q_{ij}=-2T_{ij}+g_{ij}L_{m}-2g^{ik}\frac{\delta^{2}L_{m}}{\delta g^{ij}\delta g^{lm}}
   \end{equation}
  and $f_{R}=\frac{\delta f(R,T)}{\delta R}$, $f_{T}=\frac{\delta f(R,T)}{\delta T}$ and $\nabla_{i}$ denotes the covariant derivative.\\

 Contracting on both sides of (14) we get
   \begin{equation}
   f_{R}(R,T)R+3\nabla^{k}\nabla_{k}f_{R}(R,T)-2f(R,T)=8\pi T-f_{T}(R,T)(T+Q^{i}_{i})
   \end{equation}

 Here the stress energy tensor of matter is defined by
   \begin{equation}
  T_{ij}=(p_{WD}+\rho_{WD})u_{i}u_{j}-p_{WD}g_{ij}
  \end{equation}
where $u^{i}=(1,0,0,0,0)$ is the five-velocity flow vector satisfying the relation

    \begin{equation}
        u^{i}u_{i}=1,   u^{i}\nabla_{j}u_{i}=0
     \end{equation}
   Now we can take $L_{m}=-p_{WD}$ as the Lagrangian representing the problem of wet dark fluid pressure $p_{WD}$ and energy density $\rho_{WD}$. Then making use of (15), the expression for the variation of stress-energy of matter is obtained as
   \begin{equation}
    Q_{ij}=-2T_{ij}-pg_{ij}
   \end{equation}
  Now we see that corresponding to different matter contributions from $f(R,T)$ gravity several theoretical models can be obtained. Thus depending on the matter source, there may be different class of $f(R,T)$ gravitation theory and the field equation also depends on the physical nature of the matter field through the tensor $Q_{ij}$. But here for our problem we take up the $f(R,T)$ gravitation theory of the type

   \begin{equation}
 f(R,T)=R+2f(T)
   \end{equation}
  $f(T)$ being an arbitrary function of stress-energy tensor of matter. Then making use of equation (14) the field equations for $f(R,T)$ gravitation theory take the form
   \begin{equation}
   R_{ij}-\frac{1}{2}Rg_{ij}=8\pi T_{ij}-2\dot{f}(T)T_{ij}-2\dot{f}(T)Q_{ij}+f(T)g_{ij}
   \end{equation}
  Here the dot denotes the differentiation with respect to the argument. In the case the matter source is chosen as as perfect fluid then the field equations takes the form
   \begin{equation}
   R_{ij}-\frac{1}{2}Rg_{ij}=8\pi T_{ij}+2\dot{f}(T)T_{ij}+[2p_{\omega \triangle}\dot{f}(T)+f(T)]g_{ij}
   \end{equation}

 3. \textbf{Metric and Field equations:}\\

Now for our problem we consider the metric
  \begin{equation}
  ds^{2}=dt^{2}-X^{2}(t)(dx^{2}+dy^{2}+dz^{2})-Y^{2}(t)d\phi^{2}
   \end{equation}
   Here we take the fifth coordinate $\phi$ as space-like. And we choose in such a way that
  \begin{equation}
  f(T)=kT
   \end{equation}
$k$ being constant.\\

Then making use of (17), (19), (23) and (24) in (22), and using co-moving coordinates we get the field equation in the form
\begin{equation}
 2\frac{\ddot{X}}{X}+(\frac{\dot{X}}{X})^{2}+2\frac{\dot{X}\dot{Y}}{XY}+\frac{\ddot{Y}}{Y}=(8\pi+4k)p_{WD}-k\rho_{WD}
   \end{equation}

   \begin{equation}
   3\frac{\ddot{X}}{X}+3(\frac{\dot{X}}{X})^{2}=(8\pi+4k)p_{WD}-k\rho_{WD}
   \end{equation}

   \begin{equation}
  3(\frac{\dot{X}}{X})^{2}+3\frac{\dot{X}\dot{Y}}{XY}=-(8\pi+3k)\rho_{WD}+kp_{WD}
   \end{equation}
   Here a dot denotes a differentiation with respect to time $`t'$.\\

   Now (25) and (26) give
    \begin{equation}
 \frac{\ddot{X}}{X}+2(\frac{\dot{X}}{X})^{2}-2\frac{\dot{X}\dot{Y}}{XY}-\frac{\ddot{Y}}{Y}=0
   \end{equation}
  Here we have three independent equations (25)-(27) from which four unknowns $X$, $Y$, $p_{WD}$ and $\rho_{WD}$ are to be found out. Thus we need one more relation to solve for these unknowns. Now we take up the following model.

   Model I:
     Here we assume that the expansion factor $\theta$ is proportional to the shear $\sigma^{2}$ so that
    \begin{equation}
        Y=X^{-n}
  \end{equation}
 where $n$ is a positive constant.\\

 Then from (25), (26) and (29) we get
   \begin{equation}
   X=(2-n)^{\frac{1}{2-n}}(a_{1}t+a_{2})^{\frac{1}{2-n}}
   \end{equation}
   and
    \begin{equation}
     Y=(2-n)^{-\frac{n}{2-n}}(a_{1}t+a_{2})^{-\frac{n}{2-n}}
     \end{equation}
 where $a_{1}$ and $a_{2}$ are arbitrary constants. It is observed that $2>n$ to find a viable solution.\\

 Now from (1) we have
       \begin{equation}
  p_{WD}=\omega \rho_{WD}-\omega A
   \end{equation}
   Using (32) in (26) we have
    \begin{equation}
 \rho_{WD}=[3na_{1}^{2}(\frac{1}{2-n})^{2}(a_{1}t+a_{2})^{-2}+(8\pi+4k)\omega A](8\pi \omega+4k\omega-k)^{-1}
   \end{equation}
   Then (32) gives
    \begin{equation}
 p_{WD}=\omega(8\pi \omega+4k\omega-k)^{-1}[3na_{1}^{2}(\frac{1}{2-n})^{2}(a_{1}t+a_{2})^{-2}+(8\pi+4k)\omega A]-\omega A
   \end{equation} 
And in this case we have\\

Mean Hubble parameter and scalar expansion are given by
\begin{equation}
H=(3-n)(\frac{a_{1}}{2-n})(a_{1}t+a_{2})^{-1},
\theta=(3-n)(\frac{a_{1}}{2-n})(a_{1}t+a_{2})^{-1}
\end{equation}
Deceleration parameter
\begin{equation}
q=-1+\frac{4(2-n)}{3-n}
\end{equation}
Anisotropy parameter
\begin{equation}
\triangle=0
\end{equation}

Spatial volume
\begin{equation}
V=(2-n)^{\frac{3-n}{2-n}}(a_{1}t+a_{2})^{\frac{3-n}{2-n}}
\end{equation}
Shear scalar $\sigma$  is given by
\begin{equation}
\sigma^{2}=\frac{3}{8}a_{1}^{2}(n+1)^{2}(2-n)^{-2}(a_{1}t+a_{2})^{-2}
\end{equation}

Here,
\begin{equation}
\frac{\sigma^{2}}{\theta^{2}}=\frac{3(1+n)^{2}}{8(3-n)^{2}}\neq0
\end{equation}
If $[r,s]$ are the state-finder parameters then
\begin{equation}
r=(21n^{2}-74n+65)(3-n)^{-2}
\end{equation}
and
\begin{equation}
s=\frac{3}{2}(n^{3}-n+24)(3-n)^{-1}(5n+7)^{-1}
\end{equation}
\begin{figure}
\includegraphics[height=2in]{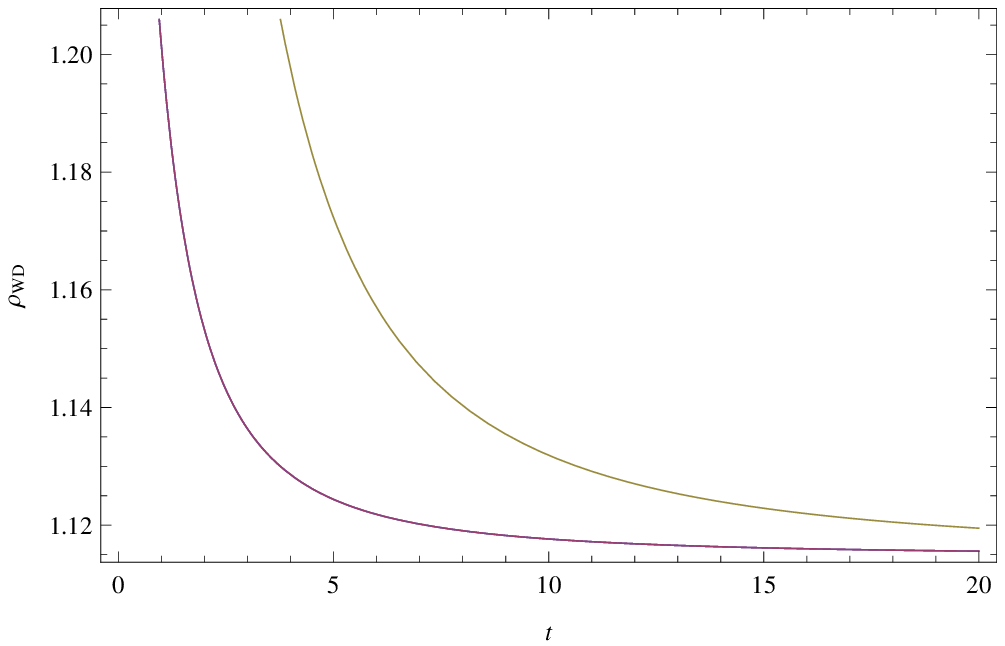}~~~~~
\includegraphics[height=2in]{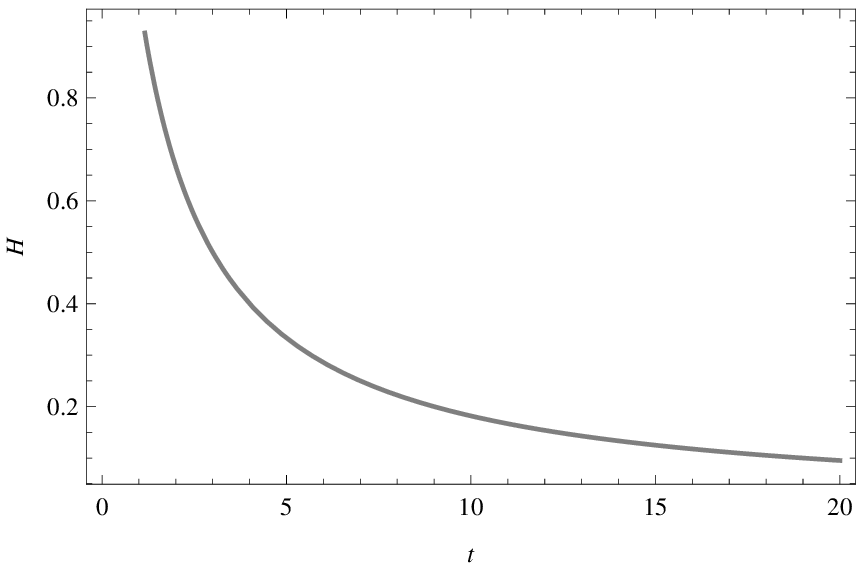}~~~~~\\

~~~~~~~~~~~~Fig.1~~~~~~~~~~~~~~~~~~~~~~~~~~~~~~~~~~~~~~~~~~~~~~~~~~~~~~~~~~~~~~~~~~~~~~~Fig.2\\

\hspace{1cm}\vspace{5mm} Fig. 1 shows variation of $\rho_{WD}$ against time $(t)$ for $\omega=\frac{1}{3}$, $k=1$, $c_{1}=1$ and $A=1$. Fig. 2 shows variation of $H$ against time $(t)$ for $n=2$ and $c_{1}=1$.

\vspace{1mm}
\vspace{1mm}

\end{figure}

\begin{figure}

\includegraphics[height=2in]{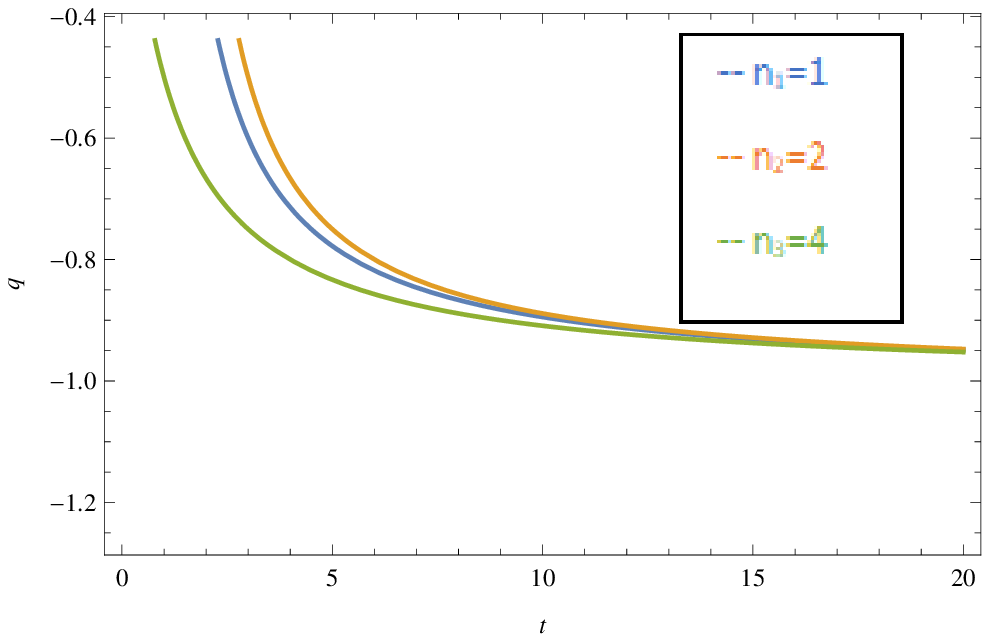}~~~~~\\

~~~~Fig.3\\

\hspace{1cm}\vspace{5mm} Fig. 3 shows variation of $q$ against time $(t)$ for $n=2$ and $c_{1}=1$.\\
\vspace{1mm}
\vspace{1mm}

\end{figure}
The plot of the density $\rho_{wd}$ with time (Fig. 1) indicates that the density is very high at early stage. But it decrease continuously with time and attains a very small positive value at late times as $t\rightarrow\infty$. The plot of $H$, shown in Fig. 2 shows that the value of Hubble parameter H always remains a positive value which shows that universe is always expanding. The plot of $q$ (Fig. 3) shows that the deceleration parameter $q$ decreases with time indicating the accelerated expansion of the universe with the passage of time.\\
Model II:
  We know that
  \begin{equation}
   \frac{\dot{V}}{V}=\frac{\frac{d}{dt}(X^{3}Y)}{X^{3}Y}=3\frac{\dot{X}}{X}+\frac{\dot{Y}}{Y}=4H
  \end{equation}

  where $V$ is the spatial volume of the Universe and $H$ is the mean Hubble parameter.\\

  Now (28) with (43) give
  \begin{equation}
  \frac{d}{dt}(\frac{\dot{X}}{X}-\frac{\dot{Y}}{Y})+(\frac{\dot{X}}{X}-\frac{\dot{Y}}{Y})\frac{\dot{V}}{V}=0
  \end{equation}
  Here in this case we consider the mixed exponential power expansion law for the spatial volume in the form
  \begin{equation}
   V=b_{0}t^{-m}e^{\frac{t^{m+1}}{m+1}}
  \end{equation}
  where $b_{0}$ and $m$ are arbitrary constants.\\

  Then we get
  \begin{equation}
   X=(b_{0}b_{2})^{\frac{1}{4}}t^{-\frac{m}{4}}exp[\frac{t^{m+1}}{4(m+1)}-\frac{b_{1}}{4b_{0}}e^{-\frac{t^{m+1}}{m+1}}]
  \end{equation}
  and
  \begin{equation}
   Y=b_{0}^{\frac{1}{4}}b_{2}^{-\frac{3}{4}}t^{-\frac{m}{4}}exp[\frac{t^{m+1}}{4(m+1)}+\frac{3b_{1}}{4b_{0}}e^{-\frac{t^{m+1}}{m+1}}]
  \end{equation}
  where $b_{0}$, $b_{1}$ and $b_{2}$ are arbitrary integration constants.\\

  Here from  (1) and (26) we get
  \begin{equation}
  \rho_{WD}=(8\pi+3k-k\omega)^{-1}[6(\frac{b_{1}}{4b_{0}}t^{m}e^{-\frac{t^{m+1}}{m+1}}+\frac{t^{m}}{4}-\frac{m}{4}t^{-1})\times(\frac{b_{1}}{4b_{0}}t^{m}e^{-\frac{t^{m+1}}{m+1}}-\frac{t^{m}}{4}+\frac{m}{4}t^{-1})-k\omega A]
  \end{equation}
  and
  \begin{equation}
  p_{WD}=\omega(8\pi+3k-k\omega)^{-1}[6(\frac{b_{1}}{4b_{0}}t^{m}e^{-\frac{t^{m+1}}{m+1}}+\frac{t^{m}}{4}-\frac{m}{4}t^{-1})\times(\frac{b_{1}}{4b_{0}}t^{m}e^{-\frac{t^{m+1}}{m+1}}-\frac{t^{m}}{4}+\frac{m}{4}t^{-1})-k\omega A]-\omega A
   \end{equation}
   In this case
   \begin{equation}
   \theta=t^{m}-\frac{m}{t}
  \end{equation}
  \begin{equation}
  H=\frac{1}{4}(t^{m}-mt^{-1})
  \end{equation}
  \begin{equation}
  q=-1-\frac{4m(t^{m+1}+1)}{(t^{m+1}-m)^{2}}
  \end{equation}

  \begin{equation}
   \sigma^{2}=\frac{3}{8}b_{0}^{-2}b_{1}^{2}t^{2m}e^{-\frac{2t^{m+1}}{m+1}}
  \end{equation}

   \begin{equation}
  \frac{\sigma^{2}}{\theta^{2}}=\frac{\frac{3}{8}b_{0}^{-2}b_{1}^{2}t^{2m+2}e^{-\frac{2t^{m+1}}{m+1}}}{(t^{m+1}-m)^{2}}
  \end{equation}

  \begin{equation}
   \triangle=3b_{0}^{-2}b_{1}^{2}(mt^{-m-1}-1)^{-2}e^{-\frac{2t^{m+1}}{m+1}}
  \end{equation}

  \begin{equation}
   r=1+\frac{12m(t^{m+1}+1)}{(t^{m+1}-m)^{2}}+\frac{16m^{2}(t^{m+1}+1)^{2}}{(t^{m+1}-m)^{4}}+\frac{16mt(t^{m+1}-m)[(m+1)t^{2m+1}+(m^{2}+3m+2)t^{m}]}{(t^{m+1}-m)^{5}}
  \end{equation}

  \begin{eqnarray}
    s&=&-\frac{1}{2}[\frac{16mt(t^{m+1}-m)\{(m+1)t^{2m+1}+(m+1)(m+2)t^{m}\}}{(t^{m+1}-m)^{3}}\nonumber\\
   &&+12m(t^{m+1}+1)+\frac{16m^{2}(t^{m+1}+1)^{2}}{(t^{m+1}-m)^{2}}](3t^{2m+2}-2mt^{m+1}+3m^{2}+4m)
  \end{eqnarray}
\begin{figure}
\includegraphics[height=2in]{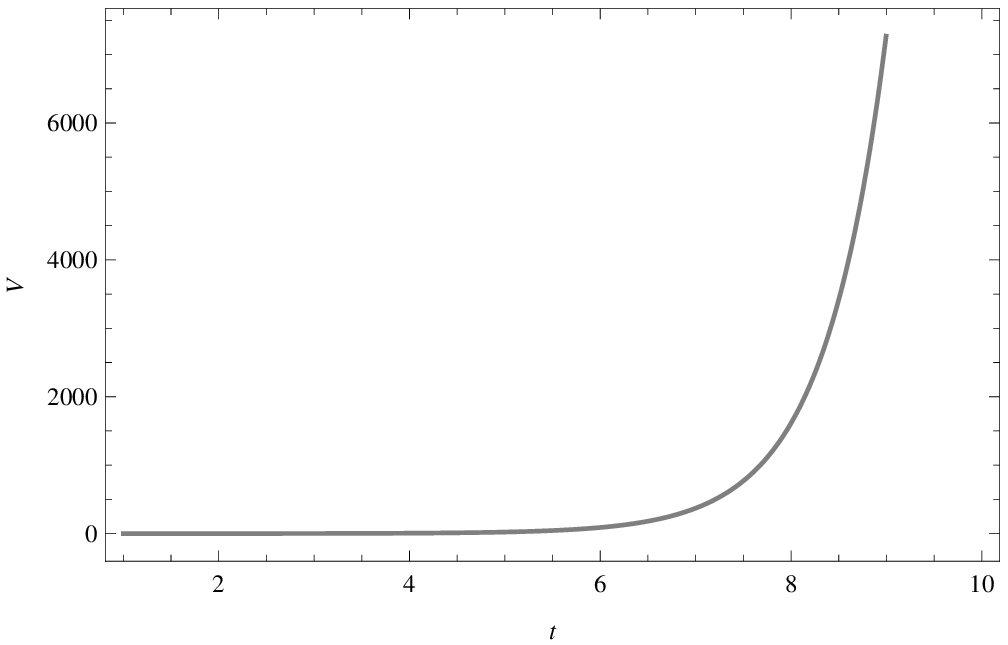}~~~~~
\includegraphics[height=2in]{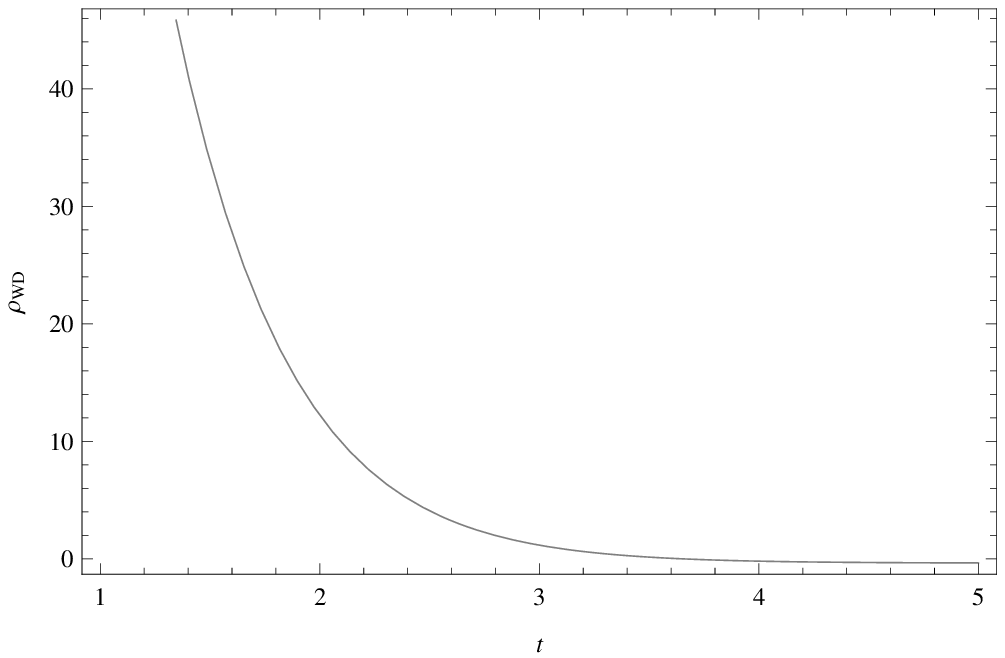}~~~~~\\

~~~~~~~~~~~~Fig.4~~~~~~~~~~~~~~~~~~~~~~~~~~~~~~~~~~~~~~~~~~~~~~~~~~~~~~~~~~~~~~~~~~~~~~~Fig.5\\

\hspace{1cm}\vspace{5mm} Fig. 4 shows variation of $V$ against time $(t)$ for $m=0.2$ and $b_{0}=0.1$. Fig. 5 shows variation of $\rho_{WD}$ against time $(t)$ for $k=1$, $b_{0}=0.1$, $b_{1}=0.1$, $b_{2}=0.1$, $m=0.2$ and $A=1$.

\vspace{1mm}
\vspace{1mm}

\end{figure}

\begin{figure}
\includegraphics[height=2in]{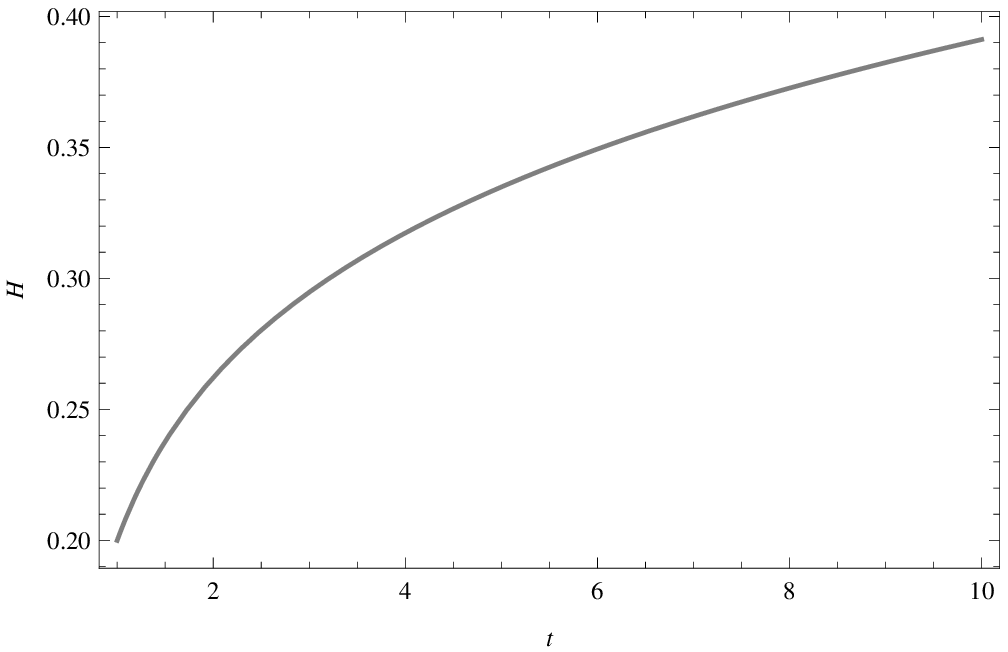}~~~~~
\includegraphics[height=2in]{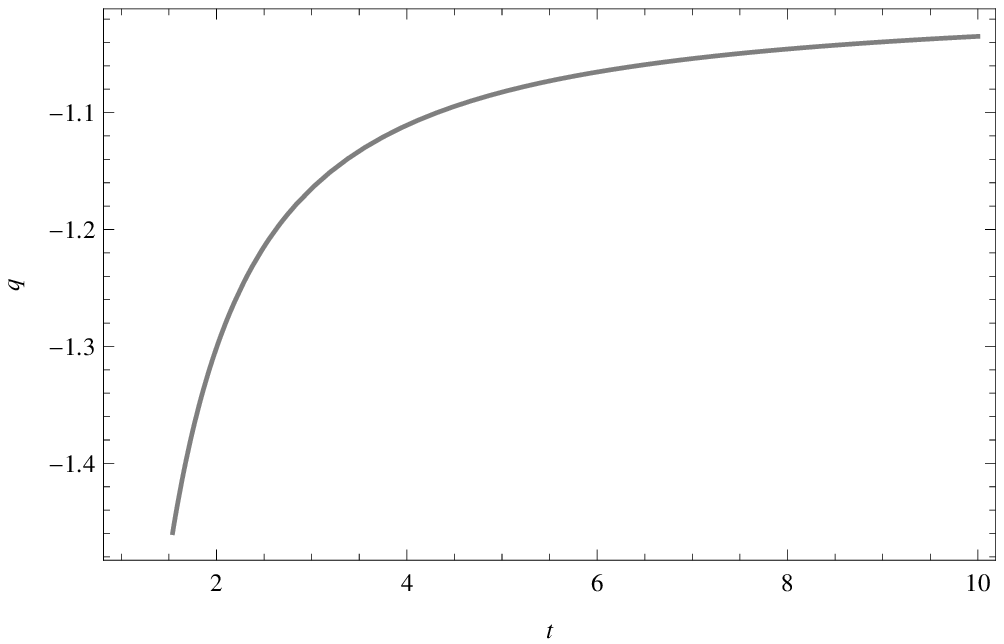}~~~~~\\

~~~~~~~~~~~~Fig.6~~~~~~~~~~~~~~~~~~~~~~~~~~~~~~~~~~~~~~~~~~~~~~~~~~~~~~~~~~~~~~~~~~~~~~~Fig.7\\

\hspace{1cm}\vspace{5mm} Fig. 6 shows variation of $H$ against time $(t)$ and Fig. 7 shows variation of $q$ against time $(t)$ for $m=0.2$.

\vspace{1mm}
\vspace{1mm}

\end{figure}

\begin{figure}
\includegraphics[height=2in]{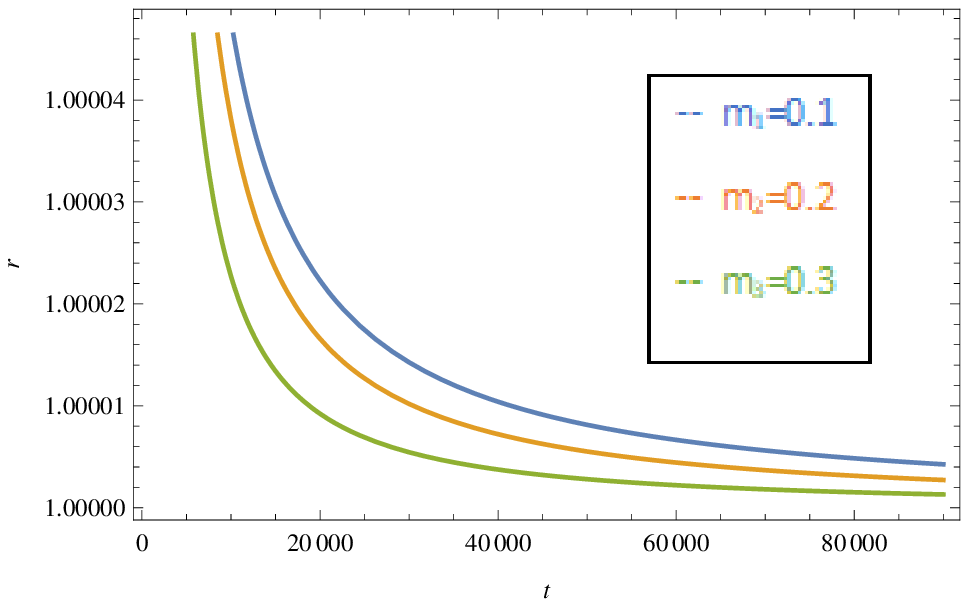}~~~~~\\

~~~~~~~~~~~~Fig.8

\hspace{1cm}\vspace{5mm} Fig. 8 shows variation of $r$ against time $(t)$.

\vspace{1mm}
\vspace{1mm}
\end{figure}
The plot of $V$ (Fig. 4) indicates that the volume is zero at $t=0$ and it increases exponentially with the passage of time. Fig. 5 shows the plot of density $(\rho_{WD})$ which shows that the density attains a negative value at late times indicating that the universe is dominated by dark energy at late times causing the late time acceleration of the universe. Fig. 6 shows the plot of Hubble parameter $H$ whose value increases wit time and is always a positive value. The plot of deceleration parameter $q$ is shown in Fig. 7 which shows tats its value is always negative which accounts for the accelerated expansion of the universe. The plot of EOS parameter $r$ (Fig. 8) shows that $r\rightarrow1$ as $t\rightarrow\infty$ indicating that the model goes to $\Lambda CDM$ model.

  Model III:\\

      Without loss of generality we can separate equation (28) into two equations namely,

  \begin{equation}
  \frac{\ddot{X}}{X}+2(\frac{\dot{X}}{X})^{2}=c_{0}^{2}
  \end{equation}
  and
  \begin{equation}
2\frac{\dot{X}\dot{Y}}{XY}+\frac{\ddot{Y}}{Y}=c_{0}^{2}
  \end{equation}
  where $c_{0}^{2}$ is the separation constant.\\

  Now from (58) we get

  \begin{equation}
  X=c_{2}e^{c_{1}(\frac{c_{0}}{\sqrt{3c_{1}}}t+1)}
  \end{equation}
  where $c_{1}$ and $c_{2}$ are constants of integration.\\

  Thus from (59) and (60) we obtain

  \begin{equation}
   Y=c_{4}e^{c_{3}(\frac{b_{0}}{\sqrt{3}b_{3}}t+1)}
  \end{equation}
  where $c_{3}$ and $c_{4}$ are arbitrary constants of integration.\\

  Here
  \begin{equation}
\rho_{WD}=\frac{\omega}{\omega+1}A+\frac{c}{b_{1}^{3(\omega+1)}b_{3}^{\omega+1}e^{(\omega+1)(\frac{4}{\sqrt{3}}b_{0}t+3b_{1}+b_{3})}}
  \end{equation}
  and
  \begin{equation}
p_{WD}=\frac{c\omega}{b_{1}^{3(\omega+1)}b_{3}^{\omega+1}e^{(\omega+1)(\frac{4}{\sqrt{3}}b_{0}t+3b_{1}+b_{3})}}-\frac{\omega}{\omega+1}A
  \end{equation}

  In this case we have

  \begin{equation}
H=\frac{c_{0}}{\sqrt{3}}
  \end{equation}
  \begin{equation}
\theta=\frac{4c_{0}}{\sqrt{3}}
  \end{equation}

  \begin{equation}
q=-1
  \end{equation}

  \begin{equation}
V=c_{2}^{3}c_{4}e^{\frac{4}{\sqrt{3}}c_{0}t+3c_{1}+c_{3}}
  \end{equation}

  \begin{equation}
\sigma^{2}=0
  \end{equation}
  \begin{equation}
\triangle=0
  \end{equation}

  \begin{equation}
\frac{\sigma^{2}}{\theta^{2}}=0
  \end{equation}

  \begin{equation}
r=1
  \end{equation}
  and
  \begin{equation}
s=0
  \end{equation}
\begin{figure}
\includegraphics[height=2in]{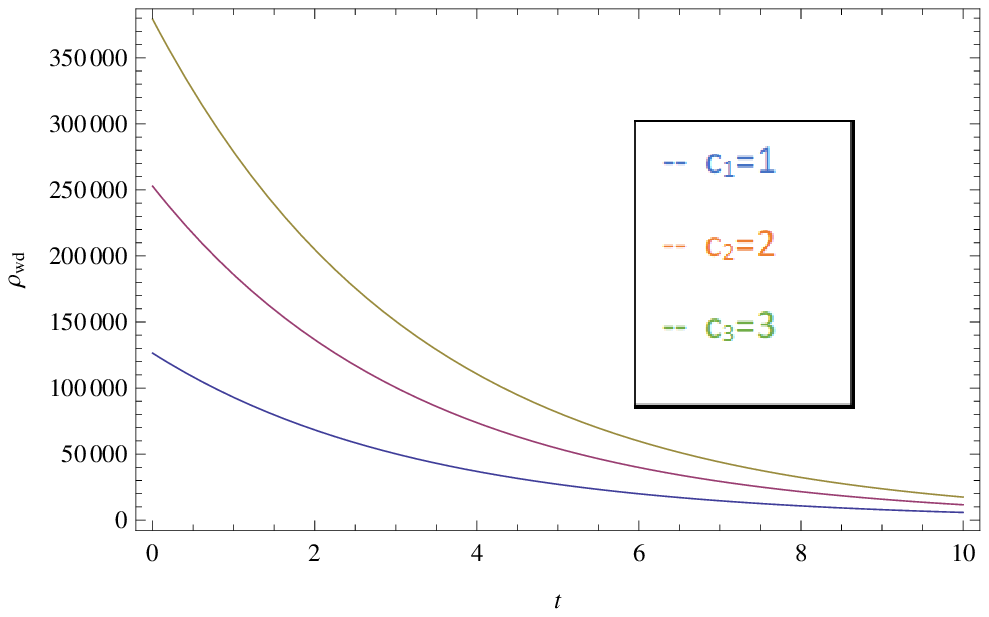}~~~~~
\includegraphics[height=2in]{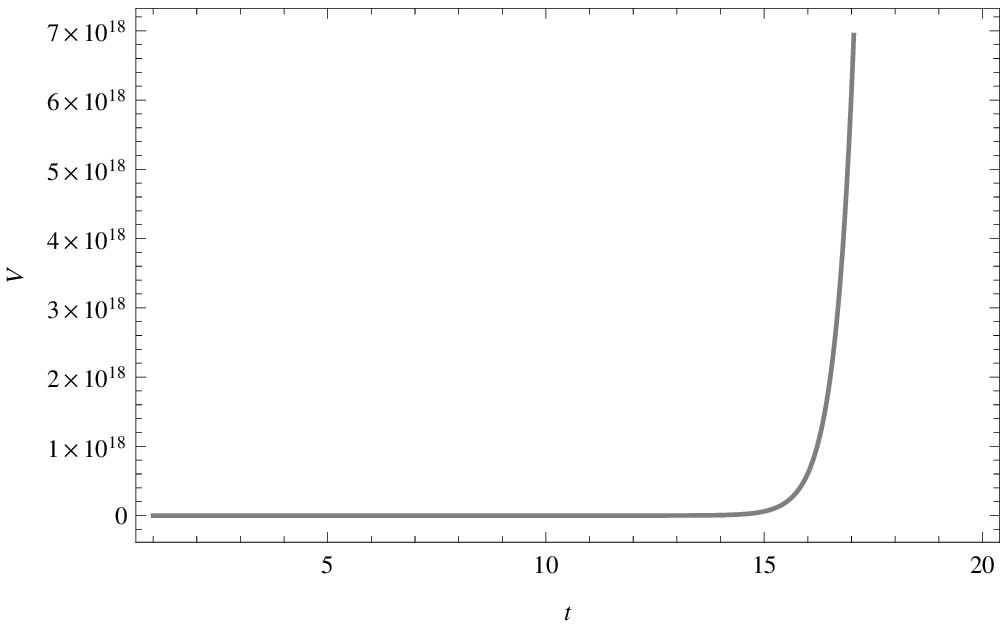}~~~~~\\

~~~~~~~~~~~~Fig.9~~~~~~~~~~~~~~~~~~~~~~~~~~~~~~~~~~~~~~~~~~~~~~~~~~~~~~~~~~~~~~~~~~~~~~~Fig.10\\

\hspace{1cm}\vspace{5mm} Fig. 9 shows variation of $\rho$ against time $(t)$  for $\omega=\frac{1}{3}$, $k=1$, $b_{0}=0.1$,$b_{1}=0.1$, $b_{2}=0.1$, $b_{3}=0.1$, $A=1$ and $c=1$. Fig. 10 shows variation of $V$ against time $(t)$ for $c_{0}=c_{1}=c_{2}=c_{3}=c_{4}=1$.
\vspace{1mm}
\vspace{1mm}
\end{figure}
The plot of the density $\rho_{wd}$ (Fig. 9) shows that the density is very high initially and there is a steady decrease with the passage of time. It tends to zero at $t\rightarrow\infty$. Fig. 10 shows the plot of $V$ which shows that the volume increases exponentially with the passage of time.

  Model IV:\\

     In this case we consider the expansion of the Universe in such a way that
  \begin{equation}
\theta=\beta t^{\alpha}
  \end{equation}
  where $\alpha$ and $\beta$ are arbitrary constants.\\

  Then we have

  \begin{equation}
3\frac{\dot{X}}{X}+\frac{\dot{Y}}{Y}=\beta t^{\alpha}
\end{equation}
Thus from (28) and (74) we have
  \begin{equation}
X=e^{\frac{\beta}{4(\alpha+1)}t^{\alpha+1}+\beta_{1}}
  \end{equation}
  and
  \begin{equation}
Y=e^{\frac{\beta}{4(\alpha+1)}t^{\alpha+1}+\beta_{0}-3\beta_{1}}
  \end{equation}
  where $\beta_{o}$ and $\beta_{1}$ are arbitrary constants.\\

Here

 \begin{equation}
\rho_{WD}=(8\pi \omega+4k\omega-k)^{-1}[\frac{3}{4}mnt^{n-1}+\frac{3}{8}m^{2}t^{2n}+(8\pi+4k)\omega A]
  \end{equation}
and
\begin{equation}
p_{WD}=\omega(8\pi \omega+4k\omega-k)^{-1}[\frac{3}{4}mnt^{n-1}+\frac{3}{8}m^{2}t^{2n}+(8\pi+4k)\omega A]-\omega A
  \end{equation}

  In this case
   \begin{equation}
H=\frac{1}{4}\beta t^{\alpha}
  \end{equation}

  \begin{equation}
q=-1-\frac{4\alpha}{\beta}t^{-\alpha-1}
  \end{equation}

  \begin{equation}
V=e^{\frac{\beta}{\alpha+1}t^{\alpha+1}+\beta_{0}}
  \end{equation}

  \begin{equation}
\sigma^{2}=0
  \end{equation}

  \begin{equation}
\triangle=0
  \end{equation}
  \begin{equation}
\frac{\sigma^{2}}{\theta^{2}}=0
  \end{equation}
  \begin{equation}
r=1+\frac{12\alpha}{\beta}t^{-\alpha-1}+\frac{16\alpha^{2}}{\beta^{2}}t^{-2\alpha-2}-\frac{16\alpha}{\beta^{2}}t^{-2\alpha-2}
  \end{equation}
  \begin{equation}
s=(6\alpha \beta t^{-\alpha-1}+8\alpha^{2}t^{-2\alpha-2}-8\alpha t^{-2\alpha-2})(3\beta^{2}+8\alpha \beta t^{-\alpha-1})
  \end{equation}
   \begin{figure}
\includegraphics[height=2in]{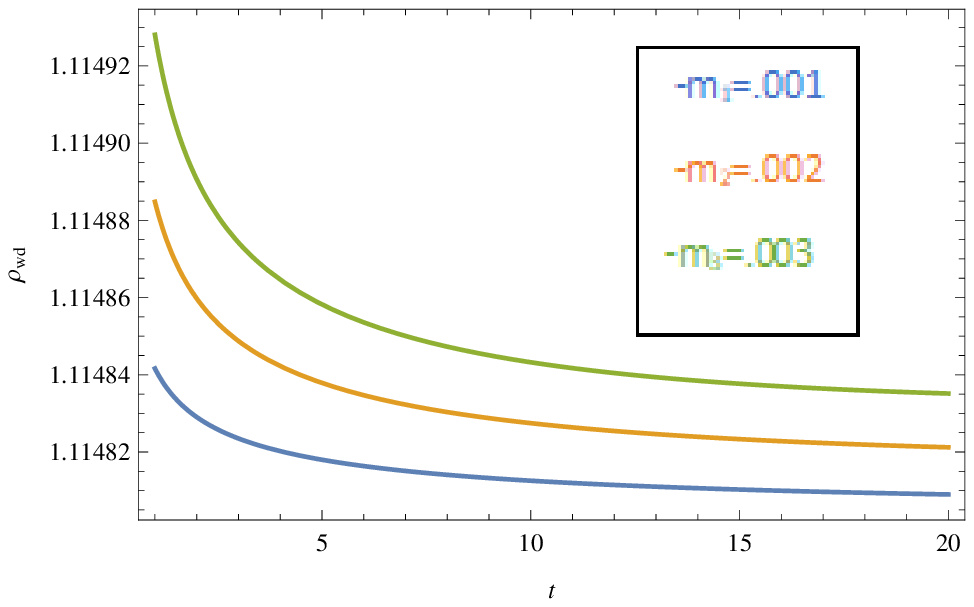}~~~~~
\includegraphics[height=2in]{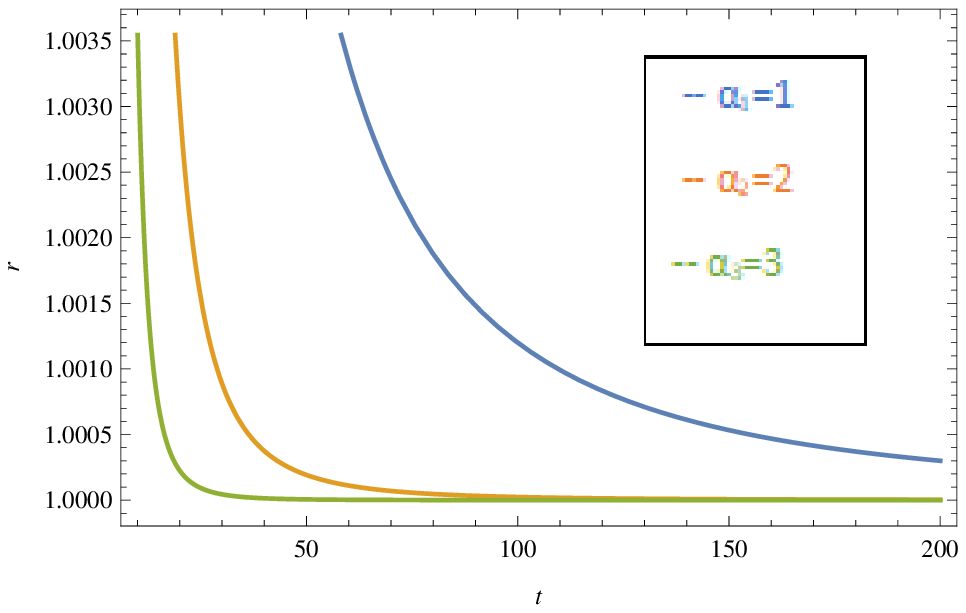}~~~~~\\

~~~~~~~~~~~~~~~~~~~~~~~Fig.11~~~~~~~~~~~~~~~~~~~~~~~~~~~~~~~~~~~~~~~~~~~~~~~~~~~~~~~~~~~~~~~~~~~~~~Fig.12~~\\

\hspace{1cm}\vspace{5mm} Fig. 11 shows variation of $\rho_{WD}$ against time $(t)$  for $\omega=\frac{1}{3}$, $k=1$ and $n=1$ and Fig. 12 shows variation of $r$ against time $(t)$ for  $\beta=1$.
\vspace{1mm}
\vspace{1mm}

\end{figure}

  \begin{figure}

\includegraphics[height=2in]{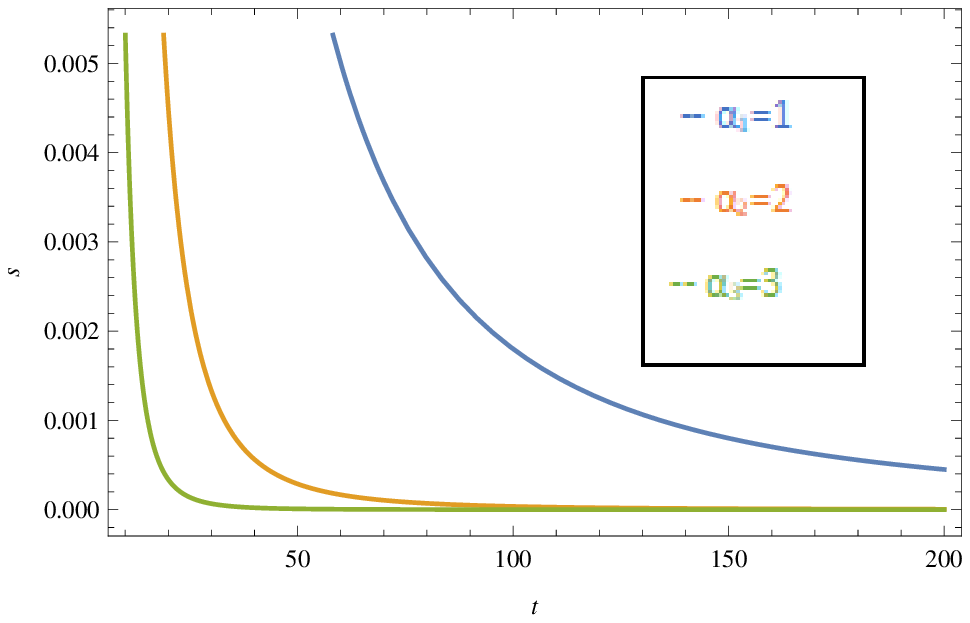}~~~~~\\

~~~~~~Fig.13\\

\hspace{1cm}\vspace{5mm} Fig. 13 shows variation of $s$ against time $(t)$ for  $\beta=1$\\.
\vspace{1mm}
\vspace{1mm}

\end{figure}

The plot of $\rho_{wd}$ (Fig. 11) shows that density is a decreasing function of time. Fig. 12 and Fig. 13 shows the graph of EOS $r$ and $s$, where $r\rightarrow\infty$ and $s\rightarrow-\infty$ as $t\rightarrow 0$  shows that it starts from asymptotic static era and then going to $\Lambda CDM$ model as $r\rightarrow 1$ and $s\rightarrow 0$ at $t\rightarrow\infty$.

  Model IV(a):\\

 Here we take as a special case, $\alpha=1$, $\beta=1$ in model IV.\\

       Then
       \begin{equation}
3\frac{\dot{X}}{X}+\frac{\dot{Y}}{Y}=t^{-1}
  \end{equation}
  Now making use of (87) in (28) we have
  \begin{equation}
\frac{\ddot{X}}{\dot{X}}-\frac{\dot{X}}{X}+\frac{1}{t}=0
  \end{equation}
  which gives
  \begin{equation}
X=d_{1}t^{d_{0}}
  \end{equation}
  where $d_{0}$ and $d_{1}$ are arbitrary constants.\\

  Now making use of (89) in (87) we have
  \begin{equation}
Y=\frac{d_{2}}{d_{1}^{3}}t^{1-3d_{0}}
  \end{equation}
  where $d_{2}$ is an arbitrary constant of integration.\\

  Here from (1) and (26) we have

  \begin{equation}
\rho_{WD}=[3d_{0}(2d_{0}-1)t^{-2}+(8\pi+4k)\omega A](8\pi \omega+4k\omega-k)^{-1}
  \end{equation}
  and
  \begin{equation}
p_{WD}=(8\pi \omega+4k\omega-k)^{-1}[3d_{0}(2d_{0}-1)t^{-2}+(8\pi+4k)\omega A]-\omega A
  \end{equation}

  In this case
  \begin{equation}
H=\frac{1}{4}t^{-1}
  \end{equation}

  \begin{equation}
\theta=t^{-1}
  \end{equation}

  \begin{equation}
V=d_{2}t
  \end{equation}

  \begin{equation}
\sigma^{2}=\frac{1}{2}(12d_{0}^{2}-6d_{0}+\frac{3}{4})t^{-2}
  \end{equation}

  \begin{equation}
\triangle=0
  \end{equation}

  \begin{equation}
q=3
  \end{equation}

  \begin{equation}
\frac{\sigma^{2}}{\theta^{2}}=\frac{1}{2}(12d_{0}^{2}-6d_{0})\neq0
  \end{equation}

  \begin{equation}
r=21
  \end{equation}

  \begin{equation}
  s=2
  \end{equation}
  \begin{figure}
\includegraphics[height=2in]{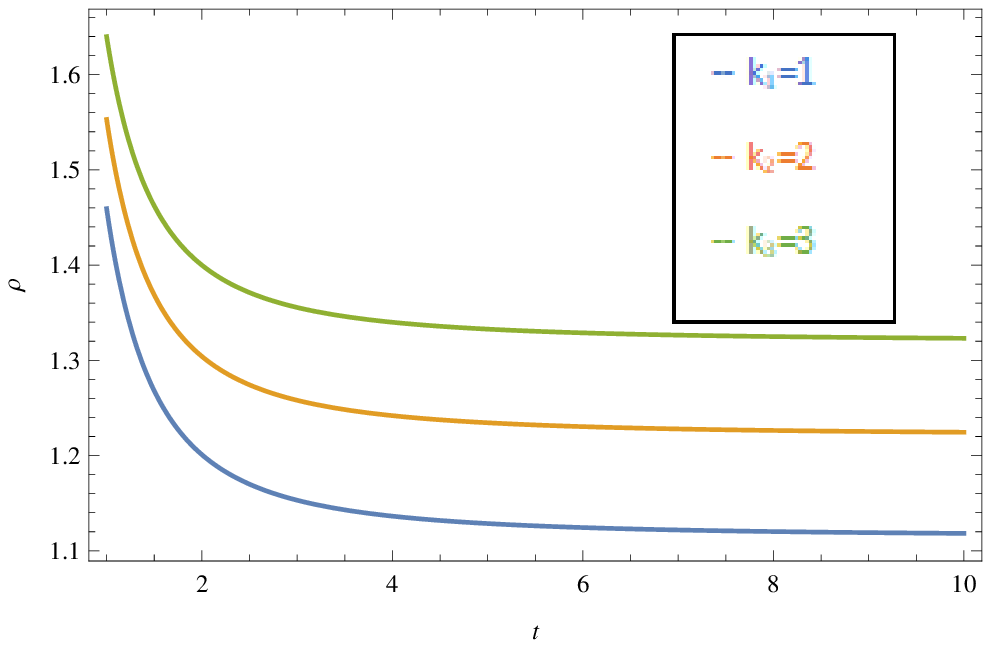}~~~~~
\includegraphics[height=2in]{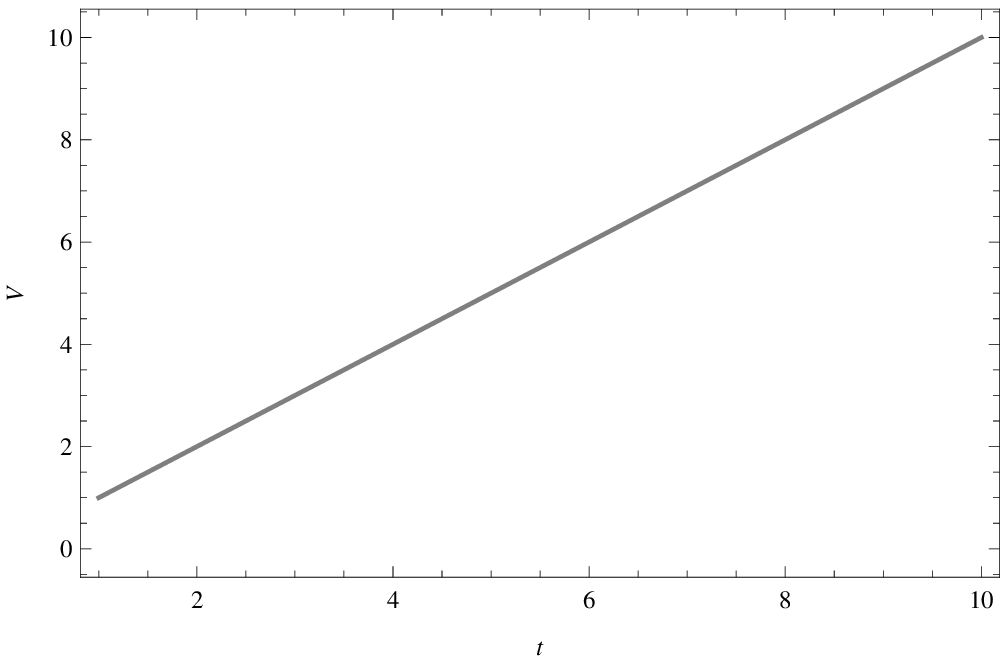}~~~~~\\

~~~~~~~~~Fig.14~~~~~~~~~~~~~~~~~~~~~~~~~~~~~~~~~~~~~~~~~~~~~~~~~~~~~~~~~~~~~~~~~~~Fig.15~~\\

\hspace{1cm}\vspace{5mm} Fig. 14 shows variation of $\rho_{WD}$ against time $(t)$  for $\omega=\frac{1}{3}$, $d_{0}=1$ and $A=1$ and Fig. 15 shows variation of $V$ against time $(t)$ for  $d_{2}=1$.\\
\vspace{1mm}
\vspace{1mm}
\end{figure}

Fig. 14 shows the graph of density $\rho_{WD}$ which indicates that the density is extremely large initially and it decrease with the passage of time and becomes negligible at late times. The plot of $V$ (Fig. 15) shows that the volume increase with time and becomes extremely large at late times.

4 {\bf Conclusions:}\\

In model I the model Universe is seen to be undergoing accelerated expansion, and it is also found that as the value of $X$ increases the value of $Y$ decreases in the inverse of the multiple of $X$, and thus as the Universe expands in an accelerated way, that is as the value of $X$ (the length of the three-dimensional space coordinates) increase rapidly the value of $Y$ (the length of the fifth dimensional coordinate) decreases more rapidly which is compatible with its contraction at a high rate so that it is invisible now which is in agreement with the present scenario of the Universe. It is seen that though this model Universe is seen to be anisotropic at the beginning of the evolution of the Universe, the anisotropy will gradually lose so that it becomes isotropic at late times as the anisotropy parameter $\triangle=0$ for this model. Here the energy density of the fluid is very high the pressure is very strong at $t=0$, thus it indicates a big bang at the beginning of this epoch. But there is no singularity as $t\rightarrow \infty$ since the pressure and the density have definite value at this instant, which indicates that there may be a beginning of the evolution of a new another era with a bounce. The energy density and the pressure of the dark energy in the form of wet dark fluid coupled with $f(R,T)$ gravitation are found to decrease with time. And it is seen that there is place for dark energy even also in $f(R,T)$ theory of gravitation; the matter content, particularly dark energy has its role in contributing to the accelerated expansion of the Universe aside from and over and above the geometric contribution to matter due to the $f(R,T)$ theory, which is the  main cause of producing accelerating expansion of the Universe according to this theory. Since this model is found to undergo inflation as well as late time accelerating expansion, it is a Universe which can explain nicely the evolution of the early Universe as well as the late Universe as it is now.

 In model II the model we obtain comes out to be a Universe with accelerating expansion. Here we see that this Universe is anisotropic in its early days, but loses gradually its anisotropy as the anisotropy parameters $\triangle$ tends to zero with the value of $t$ becoming larger which means that at late times it (the Universe) becomes isotropic. The fluid pressure is found to be negative if the parameter $\omega$ happens to be negative which is one of the main characteristics of the fluid to be dark energy. There is a singularity at $t=0$ and our model seems to begin with a big bang. At $t=m^{\frac{1}{m+1}}$ the expansion is seen to stop momentarily which may be taken as an event to bounce. The spatial volume is seen to be exceptionally large at $t=0$ which bears testimony to the big bang. At $t=0$ we see that the state-finder parameters $[r,s]$ are such that $r\rightarrow \infty$  and $s\rightarrow -\infty$ which indicates that our model starts from asymptotic static era, and it goes to $\Lambda CDM$ model $[r=1, s=0]$ at later era. Here it is found that the geometric contribution of matter due to $f(R,T)$ gravity cause the accelerated expansion of the Universe, and also this expansion is enhanced by the presence of wet dark fluid playing the role of dark energy. In this model the energy density and fluid pressure are found to become infinite as $t\rightarrow-\infty$, hence it may be concluded that there was another epoch of this Universe from $t=-\infty$ to $t=0$ starting with high energy density and a strong pressure at the beginning.

 In model III the model obtained is seen to be a Universe with accelerating expansion. The spatial volume is increasing here exponentially. This model Universe happens to represent at late times $\Lambda CDM$ Universe as the state-finder take the values $[r=1, s=0]$ as $t\rightarrow0$. The energy density and the fluid pressure both do not vanish at $t=0$ as well as at $t\rightarrow\infty$. Thus this model may be either oscillatory or cyclic type of Universe. Since spatial volume is very much related to the scale factors of the Universe, from the expression of the energy density and fluid pressure of the wet dark fluid it is seen that the dark energy contributes here also to the accelerating expansion of the Universe apart from enhancing of expansion through geometric contribution of matter due to $f(R,T)$ gravitation. This model is found to be isotropic which is in agreement with the present day scenario of the Universe. No singularity  is seen in this Universe. The pressure and the density are found to be very high as $t\rightarrow-\infty$, thus it seems that there was an era of this model Universe in the remote past beginning with a big bang, before the beginning of the present era. The model Universe obtained in model IV is also a Universe with accelerating expansion. At $t=0$ there is no expansion, but as $t\rightarrow0$ the accelerating expansion rate increases enormously with high degree of acceleration, thus there is a fear that there may be a singularity event at this moment of time and perhaps may be a big rip singularity. At $t=0$ the spatial volume of the Universe takes a definite value which means almost that the Universe existed before $t=0$. Thus it is seen that the Universe is undergoing accelerating expansion due to the geometric contribution to matter. Here from the expression of the density it is seen that the matter content is partly from the dark energy and partly from the geometric contribution due to $f(R,T)$ gravitation. Because of these factors the pressure and the density seem to exist even at $t=0$. In this case the density becomes zero when the dark energy parameter $\omega$ and $f(R,T)$ gravitation theory parameter $k$ are related in the form $k=\frac{8\pi \omega}{1-4\omega}$ or $\omega=\frac{k}{8\pi+4k}$ which seems that under these conditions the contribution to matter due to dark energy and due to $f(R,T)$ gravitation theory are equal and opposite to each other. But at this moment the pressure remains and is equal to $-\omega A$ which shows that the contribution due to the dark energy part dominate the other part. Here the non-geodesic motion of test particles are resulted from the coupling of matter with geometry in this gravitational theory and an extra accelerated motion is produced. This Universe is seen to start from asymptotic static era as state-finder parameters $[r,s]$ are found in the form $r\rightarrow\infty$ and $s\rightarrow-\infty$ as $t\rightarrow0$ and then going to $\Lambda CDM$ model as we find $r=1$ and $s=0$ as $t\rightarrow\infty$. In this model the spatial volume is also increasing at a high rate bearing testimony to the happening of a big rip singularity.

 In model IV(a) we get our model as an expanding Universe. Here the scale factor $X$ is found to increase with time whereas the scale factor $Y$ is found to decrease at a high rate so that at late times it is so small that it is invisible as it is at the present time. Here the deceleration parameter is found to be positive and therefore our Universe is decelerating initially but will accelerate at late time because of ``cosmic re-collapse" (Nojiri and Odinstov, 2003). Thus our model will be a Universe with accelerating expansion in unison with the present era. Here the volumetric expansion is linear, it may be because of inflation due to high rate of expansion at present. The energy density and the fluid pressure are so high at $t=0$, thus it seems that there is a big bang at this moment. The density and the pressure do not vanish neither at $t\rightarrow\infty$ nor at $t\leftarrow-\infty$, their values being solely dependent on the parameters $k$, $\omega$, $A$ at these moments. Thus it may be concluded that this type of Universe must be either an oscillatory or a cyclic type. Here it is seen that as $\frac{\sigma^{2}}{\theta^{2}}$ is not zero, our model Universe is anisotropic at the beginning, but as the anisotropy parameters $\triangle=0$ it (the Universe) gradually loses the anisotropic-ness and becomes isotropic at late time which is in agreement with the scenario of our actual Universe. Here the shear is found to decrease with time until it becomes $0$ as $t\rightarrow\infty$. In this model there is a singularity at $t=0$ with very high density and pressure, thus this model Universe started its evolution with a big bang in the beginning.


\begin{thebibliography}\\
\bibitem{} S. Perlmutter, G. Aldering, M.D. Valle, et al. Nature,391, 51 (1998). doi:10.1038/34124.\\
\bibitem{} A.G. Riess, A.V. Filippenko, P. Challis, et al. Astron. J.116, 1009 (1998). doi:10.1086/300499.\\
\bibitem{} B. Ratra, P.J.E. Peebles, Phys. Rev. D 37, 3406 (1988).\\
\bibitem{} C. Armendariz-Picon, V. Mukhanov, P.J. Steinhardt, Phys. Rev. D 63, 103510 (2001).\\
\bibitem{} S.M. Carroll, M. Hoffman, M. Trodden, Phys. Rev. D 68, 023509 (2003).\\

\bibitem{} S.M. Carrol, V. Duvvuri, M. Trodden, and M.S. Turner. Phys. Rev. D, 70,043528 (2004). doi:10.1103/PhysRevD.70.043528.\\
\bibitem{} S. Capozziello, S. Nojiri, S.D. Odintsov, and A. Troisi. Phys. Lett. B,639, 135(2006). doi:10.1016/j.physletb.2006.06.034.\\
\bibitem{} S. Nojiri and S.D. Odintsov. Phys. Rev. D, Part. Fields, 74, 086005 (2006).doi:10.1103/PhysRevD.74.086005.\\
\bibitem{} M. Sharif and S. Arif. Astrophys. Space Sci.342, 237 (2012). doi:10.1007/s10509-012-1150-2.\\

\bibitem{} S. Nojiri, S.D. Odintsov, and P.V. Tretyakov. Prog. Theor. Phys. Suppl.172,81(2008). doi:10.1143/PTPS.172.81.\\
\bibitem{} N.M. Garcia, F.S.N. Lobo, and J.P. Mimoso. J. Phys. Conf. Ser.314, 012056(2011). doi:10.1088/1742-6596/314/1/012056.\\
\bibitem{} K. Bamba, S. Capozziello, S. Nojiri, and S.D. Odintsov. Astrophys. Space Sci.342, 155 (2012). doi:10.1007/s10509-012-1181-8.\\
\bibitem{} M. Jamil, D. Momeni, and R. Myrzakulov. Eur. Phys. J. C,72, 2137 (2012).doi:10.1140/epjc/s10052-012-2137-4.\\

\bibitem{} T. Harko, F.S.N. Lobo, S. Nojiri, and S.D. Odintsov. Phys. Rev. D, Part. Fields,84, 024020 (2011). doi:10.1103/PhysRevD.84.024020.\\
\bibitem{} B.C. Paul, P.S. Debnath, and S. Ghose. Phys. Rev. D,79, 083534 (2009). doi:10.1103/PhysRevD.79.083534.\\
\bibitem{} M. Sharif and M.F. Shamir. Class. Quantum Gravity,26, 235020 (2009). doi:10.1088/0264-9381/26/23/235020.\\
\bibitem{} M. Sharif and M.F. Shamir. Mod. Phys. Lett. A,25, 1281 (2010). doi:10.1142/S0217732310032536.\\
\bibitem{} M.F. Shamir. Astrophys. Space Sci.330, 183 (2010). doi:10.1007/s10509-010-0371-5.\\
\bibitem{} Shamir, M.F., Bhatti, A.H.: Can. J. Phys.06, 90 (2012)\\
\bibitem{} Brans, C.H., Dicke, R.H.: Phys. Rev.124, 925 (1961)\\
\bibitem{} Reddy, D.R.K., Naidu, R.L., Satyanarayana, B.: Int. J. Theor. Phys.51,3045 (2012a)\\
\bibitem{} Saez, D., Ballester, V.J.: Phys. Lett. A113, 467 (1986)\\
\bibitem{} Reddy, D.R.K., Naidu, R.L., Satyanarayana, B.: Int. J. Theor. Phys.51,3222 (2012b)\\
\bibitem{} Rao, V.U.M., Neelima, D.: Int. Sch. Res. Not. (2013). doi:10.1155/2013/174741\\
\bibitem{} Reddy, D.R.K., Satyanarayana, B., Naidu, R.L.: Astrophys. Space Sci.339, 401 (2012c)\\
\bibitem{} Sahoo, P.K., Mishra, B.: Can. J. Phys.92, 1062 (2014)\\
\bibitem{} Reddy, D.R.K., Santhi Kumar, R., Pradeep Kumar, T.V.: Int. J. Theor.Phys.52, 239 (2013)\\
\bibitem{} Rao, V.U.M., Suryanarayana, G.: Prespacetime J.5, 1389 (2014)\\




\end{thebibliography}
\end{document}